\documentclass[%
 reprint,
 amsmath,amssymb,
 aps,
]{revtex4-2}

\usepackage{graphicx}
\usepackage{dcolumn}
\usepackage{mathrsfs}
\usepackage{bm}

\begin{document}

\preprint{APS/123-QED}

\title{Piezoelectrically driven diamond phononic nanocavity \\by phonon-matching scheme for quantum applications}

\author{Michele Diego$^1$}
\email{diego@iis.u-tokyo.ac.jp}
\author{Byunggi Kim$^1$}
\author{Matteo Pirro$^1$}
\author{Sebastian Volz$^{1,2}$}
\author{Masahiro Nomura$^1$}
\email{nomura@iis.u-tokyo.ac.jp}
\affiliation{$^1$Institute of Industrial Science, The University of Tokyo, Tokyo 153-8505, Japan}
\affiliation{$^2$Laboratory for Integrated Micro and Mechatronic Systems, CNRS-IIS UMI 2820, The University of Tokyo, Tokyo 153-8505, Japan}%
%


\date{\today}

\begin{abstract}
Efficiently exciting and controlling phonons in diamond nanoresonators represents a fundamental challenge for quantum applications. Here, we theoretically demonstrate the possibility to excite mechanical modes within a double-hybrid-cavity (DHC), formed by adjoining to a diamond cavity a second cavity, made of aluminum nitride. 
The latter is piezoelectric and serves as a microwave-to-phonon transducer, activating mechanical modes in the entire DHC.
We show the process to match the cavities phononic properties, making them work coordinately in the DHC and obtaining a well confined mode. 
In the diamond part of the cavity, this mode replicates the fundamental mode of the individual diamond cavity, showing that the piezoelectric transducer doesn't alter the diamond individual fundamental mode. In the piezoelectric part, the strong confinement of stress and electric field results in a high piezoelectric coupling rate, demonstrating the effectiveness of a phononic cavity as a transducer. The study is contextualized in the framework of a quantum networking application, where the DHC serves as a spin qubit, exploiting the spin-mechanical coupling within diamond color centers.
\end{abstract}

\maketitle


\section{\label{sec:level1}Introduction}
Diamond structures embedding color centers \cite{iwasaki2015germanium, castelletto2020silicon, westerhausen2020controlled, tchernij2021spectral} are at the forefront of quantum remote communication \cite{dolde2013room, bernien2013heralded, ruf2021quantum, sekiguchi2021geometric}.
In particular, diamond mechanical nanoresonators incorporating negatively charged nitrogen-vacancy (NV) color centers \cite{jelezko2004observation, bar2013solid, nagata2018universal} are emerging as potential candidates for quantum networking \cite{shandilya2021optomechanical}. 
Indeed, two-levels spin transitions can be controlled in diamond NV centers by exploiting their sensitivity to mechanical stress \cite{d2011quantum, kolkowitz2012coherent, macquarrie2013mechanical, teissier2014strain, yeo2014strain, lekavicius2019diamond}, making them ideal qubits. 
Diamond qubits can therefore serve as nodes in a quantum network, where the information on the spin state of each qubit can be read out optically \cite{golter2014optically, aharonovich2016solid, gao2022narrow} and transmitted over long distances to other nodes, thereby interconnecting the network. More specifically, each node should incorporate a phonon source at the transition frequency of NV centers and a photonic waveguide for readout, connected to a telecommunication optical fiber.

One of the main challenges is then how to efficiently excite and control the mechanical modes in diamond structures.
Recent advances in hybrid devices, integrating diamond and piezoelectric structures, offer an efficient solution to embed NV centers and directly excite mechanical modes through the inverse piezoelectric effect \cite{honl2022microwave}. Raniwala \textit{et al.} proposed an electromechanical device where a diamond resonator is excited by a piezoelectric transducer \cite{raniwala2023piezoelectric}. Ding \textit{et al.} demonstrated the possibility to excite mechanical modes in a diamond waveguide by surface acoustic waves \cite{ding2024integrated}. Another strategy is offered by our recent work \cite{kim2023diamond}, where a piezoelectric pad is attached on top of a diamond optomechanical cavity, and works as a microwave transducer to excite phonons.

In this type of structures, the aim is to excite the diamond resonator without compromising its fundamental resonant mode. 
Generally, indeed, the closer and larger the piezoelectric transducer to the resonant mode localized in diamond, the stronger the excitation. However, the mass of the transducer alters the resonant mode's shape, frequency and coupling with color centers. This trade-off limits the piezoelectric coupling rate, which quantifies the effectiveness of piezoelectric excitation and thus the performance of the network node. In current devices, this coupling rate falls within the MHz range \cite{kurokawa2022remote, kim2023diamond}. Ideally, a perfect transducer would achieve a high piezoelectric coupling rate without affecting the diamond resonant mode features.

Here, we theoretically introduce a structure approaching the ideal case, with an efficient piezoelectric transducer attached to the localized region of the diamond mode, but without altering its shape or the coupling with the NV centres. Such a configuration achieves a piezoelectric coupling rate four orders of magnitude higher than for the cavity with the piezoelectric pad \cite{kim2023diamond}, when the two structures have similar sizes. The design is realized by adjoining two quasi one-dimensional patterned phononic cavities, respectively made of diamond and aluminium nitride (AlN), thus forming a double-hybrid-cavity (DHC). 

The challenge lies in the non-trivial integration of two cavities made of different materials to form a coherent DHC. 
Phononic cavities, indeed, rely on two key characteristics. Firstly, a periodic unit cell that leads to destructive interference of acoustic waves \cite{anufriev2021review, nomura2022review}, resulting in the formation of bandgaps in the phonon dispersion \cite{safavi2014optomechanical, raniwala2022spin}. Secondly, geometrical defects that locally modify the phonon dispersion and cause the localization of mechanical modes at frequencies within the unperturbed bandgap \cite{maccabe2020nano}. In this work, we show the process to achieve both these two key elements for a DHC. Then, we quantitatively discuss the advantages provided by the DHC.

We start with the customization of the phonon dispersions of the individual diamond and piezoelectric unit cells, aiming to open a bandgap in the hybrid diamond/piezoelectric double unit cell phonon dispersion. Then, we utilize a homemade genetic algorithm to optimize the DHC defects, effectively confining its fundamental resonant mode within the newly formed bandgap.

The optimized resonant mode replicates, in the diamond part, the fundamental mode of the individual diamond cavity (DC), thus assuring similar performance in terms of spin-mechanical coupling with the color centers. This demonstrates that the piezoelectric cavity (PC) can be used as a transducer without altering the diamond resonant mode.
Within the PC, the overlap between the confined stress and electric fields leads to the high piezoelectric coupling rate, demonstrating its effectiveness as a transducer.\\
\begin{figure*}[thb]
\centering
\includegraphics[width=0.975\textwidth]{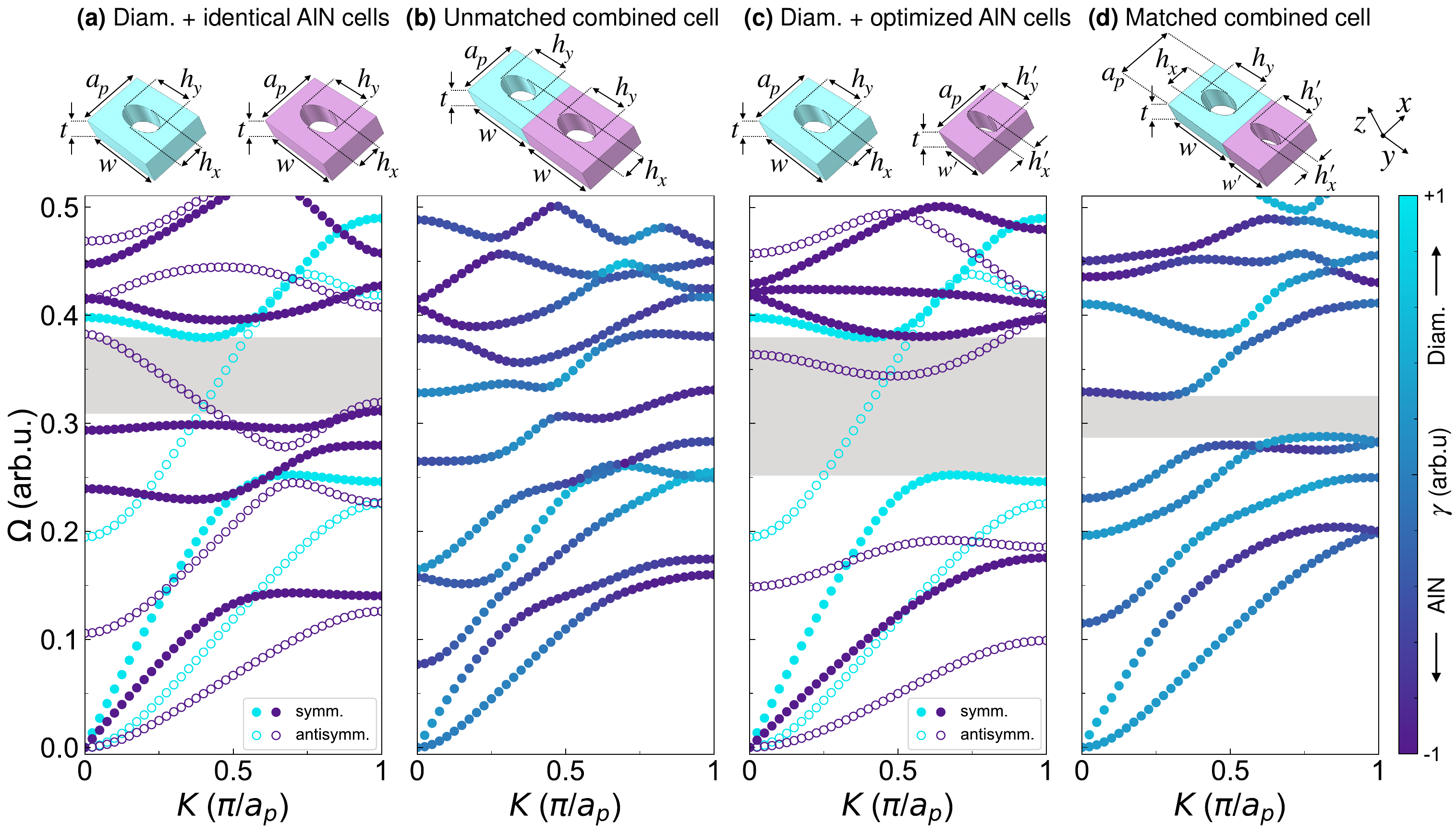}
\caption{Different diamond, AlN and hybrid unit cells and relative $z$-symmetric phonon dispersions. (a) diamond unit cell (light blue) with thickness $t$, periodicity $a_p$, width $w$ and patterned elliptical hole of axis $h_x$ and $h_y$. AlN identical cell (light purple) having the same parameters as the diamond cell. Phonon dispersion of the diamond unit cell (light blue circles) and identical AlN unit cell (purple circles). Full and empty circles represent, respectively, $y$-symmetric and $y$-antisymmetric modes. The intersection of diamond/AlN symmetric bandgaps is highlighted in grey.
(b) identical (unmatched) double hybrid diamond/AlN unit cell and relative phonon dispersion. Each point is color-weighted according to $\gamma = (U_{mec}^{(d)}-U_{mec}^{(p)})/(U_{mec}^{(d)}+U_{mec}^{(p)})$. The phonon dispersion doesn't present any bandgap.
(c) unmodified diamond unit cell (light blue) and AlN optimized unit cell (light purple) with width $w'$ and patterned elliptical hole with axis $h_x'$ and $h_y'$. Phonon dispersion of the diamond unit cell (light blue circles) and optimized AlN unit cell (purple circles). The color/symbol code is the same as in (a).
(d) optimized (matched) double hybrid diamond/AlN unit cell and relative phonon dispersion. The color code is the same as in (b). In this case, the phonon dispersion presents a bandgap.
}
\label{fig:1}
\end{figure*}
\section{Results and discussion}
\subsection{Tuning the phonon dispersion}
To design the DHC with a well-confined resonant mode, we first need to open a bandgap in the phonon dispersion associated to the double hybrid diamond/piezoelectric unit cell. Here, we show the strategy to obtain this.

We start by taking a quasi one-dimensional diamond nanocrystal with geometrical axes orientated along the diamond lattice principal axes. The nanocrystal is formed by the repetition along the $x$-coordinate of the unit cell sketched in Fig. \ref{fig:1}(a) (top left, in light blue), with periodicity $a_p$ and thickness $t=0.37a_p$.
The unit cell has a rectangular cross section of width $w=1.43a_p$, with a patterned elliptical hole at the center, whose axes are $h_x=0.37a_p$ and $h_y= 0.78a_p$ (more details about all the structures can be found in the Appendix \ref{app:A}). 
In the same panel, we show with light blue circles the $z$-symmetric phonon dispersion associated to such a diamond unit cell, plotting the reduced frequency $\Omega = \omega a_p / (2 \pi v_t)$ as a function of the wavevector $K$ ($\omega$ is the angular frequency and $v_t=12800$ m/s the sound transverse velocity in diamond). The $\alpha$-symmetry is to be intended in respect to the plane cutting in half the cell and perpendicular to the $\alpha$-axis ($\alpha=x,y,z$). All the phonon dispersion curves in Fig. \ref{fig:1} are $z$-symmetric. 
Diamond unit cell branches are further differentiated into $y$-symmetric and $y$-antisymmetric.
A $y$-symmetric bandgap is present between approximately 0.25 and 0.38.

In the same plot, purple circles indicate the phonon dispersion of an AlN unit cell, whose geometry is identical to the diamond one and is displayed at the top of the panel (in light purple). The crystal orientation is assumed to have the out-of-plane lattice axis along the $z$-coordinate and in-plane equivalent principal lattice axes along, respectively, $x$ and $y$ coordinates. This cell exhibits three $y$-symmetrical bandgaps between approximately 0.14 and 0.23, 0.27 and 0.3, and finally 0.31 and 0.4. Therefore, the main intersection between $y$-symmetrical bandgaps of the diamond and the AlN cells with identical shape lays between approximately 0.31 and 0.38. This zone is highlighted in grey for an easier visualization.

We now laterally adjoin the diamond cell to the identical AlN cell, assuming the continuity of the displacement at the interface between the two materials. The resulting cell is then a double hybrid cell, whose geometry is shown Fig. \ref{fig:1}(b) (top). Since the side of the double cell is still $a_p$, its periodic repetition generates a hybrid nanocrystal, possessing a well-defined phonon dispersion, shown as well in Fig. \ref{fig:1}(b).
The presence of different materials breaks the symmetry along the $y$-axis, therefore modes can no longer be distinguished in $y$-symmetrical and $y$-antisymmetrical. Nevertheless, a classification of the modes can be performed by examining their origins. In many cases, modes from the individual unit cells can be identified within those of the combined unit cell. These are modes that are only minorly perturbed by the lateral contact with the other material in the combined cell. As a result, their energy is predominantly localized in either the diamond or the AlN part of the cell. 
Branches formed by these modes are a product of hybridization between branches of individual unit cells. To visualize this effect, each mode in Fig. \ref{fig:1}(b) is color-weighted with $\gamma = (U_{mec}^{(d)}-U_{mec}^{(p)})/(U_{mec}^{(d)}+U_{mec}^{(p)})$, where $U_{mec}^{(d)}$ is the mechanical mode energy contained in diamond, while $U_{mec}^{(p)}$ is the one contained in AlN. For each $m$-eigenmode with reduced frequency $\Omega^{(m)}$ (and angular frequency $\omega^{(m)}=2 \pi v_t \Omega^{(m)}  / a_p$), $U_{mec}^{(\beta)}$ ($\beta=d,p$) is given by the temporal average of the mechanical energy over a complete cycle with period $\tau^{(m)}=2\pi/\omega^{(m)}$:
\begin{equation}\label{eq:mecenergy}
U_{mec}^{(\beta)}=\frac{1}{\tau^{(m)}} \int_0^{\tau^{(m)}} d\text{t} \int_{\beta} d\textbf{r} \mathscr{U}_{mec}(\textbf{r},t).
\end{equation}
$\mathscr{U}_{mec}$ is the mechanical energy density, which can be calculated from the kinetic and potential  contributions:
\begin{equation}\label{eq:energydensity}
\mathscr{U}_{mec} = \frac{1}{2}\rho v^2 + \frac{1}{2} \textbf{T} : \bm{\epsilon}, 
\end{equation}
where $\rho$, $v$ and  $\textbf{T}: \bm{\epsilon}$ are, respectively, the density, the velocity and the tensor product between the stress and the strain fields in the $m$-eigenmode. The spatial integral spans over the diamond ($\beta=d$) or the AlN ($\beta=p$) part of the unit cell. Thus, $\gamma   \rightarrow 1$ indicates modes that predominantly derive from the diamond cell modes, and $\gamma \rightarrow -1$ from the AlN cell.\\
The hybridised origin of many of the branches implies that, as a rule of thumb, the combined cell is more likely to exhibit a bandgap in frequency ranges where both individual cells have a low density of modes, as within bandgaps.
Enlarging the bandgap intersection between the individual cells, then, increases the chances of finding a bandgap in the combined cell.
In the case of the dispersion curve in Fig. \ref{fig:1}(b), no bandgap is present for the double hybrid cell, indicating that the symmetric bandgaps intersection of the individual cells is not sufficient to assure a good phonon-matching.
For this reason, we refer to the hybrid unit cell formed by identical diamond and AlN cells as the unmatched unit cell. This implies that —as we will see in detail in the following— it is not possible to localize efficiently mechanical modes in an unmatched DHC created from such a unit cell.

We now tailor the AlN unit cell, aiming to open a bandgap in the double hybrid cell phonon dispersion. This is done by tuning its individual phonon dispersion to enlarge the intersection between its bandgap and the diamond cell one. Figure \ref{fig:1}(c) shows the optimized AlN cell rendering (top right, in light purple) and its phonon dispersion. The new geometrical parameters of the cell are, respectively, $w'=0.9a_p$, $h_x'=0.22a_p$ and $h_y'=0.65a_p$. The associated phonon dispersion presents a large $y$-symmetric bandgap between approximately 0.17 and 0.38. Figure \ref{fig:1}(c) also shows again the unmodified diamond cell sketch (top left, in light blue) and its phonon dispersion (light blue circles), allowing a direct comparison with the optimized AlN one. The intersection between their $y$-symmetric bandgaps, highlighted in grey, extends in the range between approximately 0.25 and 0.38, i.e. almost twice as large as the one for identical cells in Fig. \ref{fig:1}(a).\\
The impact of such a large bandgap intersection is reflected in the phonon dispersion of the double hybrid cell formed by the adjunction of the diamond unit cell and the optimized AlN unit cell. This is shown in Fig. \ref{fig:1}(d), displaying the rendering (top) and the dispersion curve of the such unit cell. At variance from the unmatched unit cell (Fig. \ref{fig:1}(b)), here a bandgap is present between approximately 0.29 and 0.32 (highlighted in grey). For this reason, we refer to this unit cell as the matched unit cell.
Thanks to the opening of the bandgap, the matched hybrid cell is suited to form a matched DHC capable of confining mechanical modes.\\
\begin{figure}[t]
\centering
\includegraphics[width=0.49\textwidth]{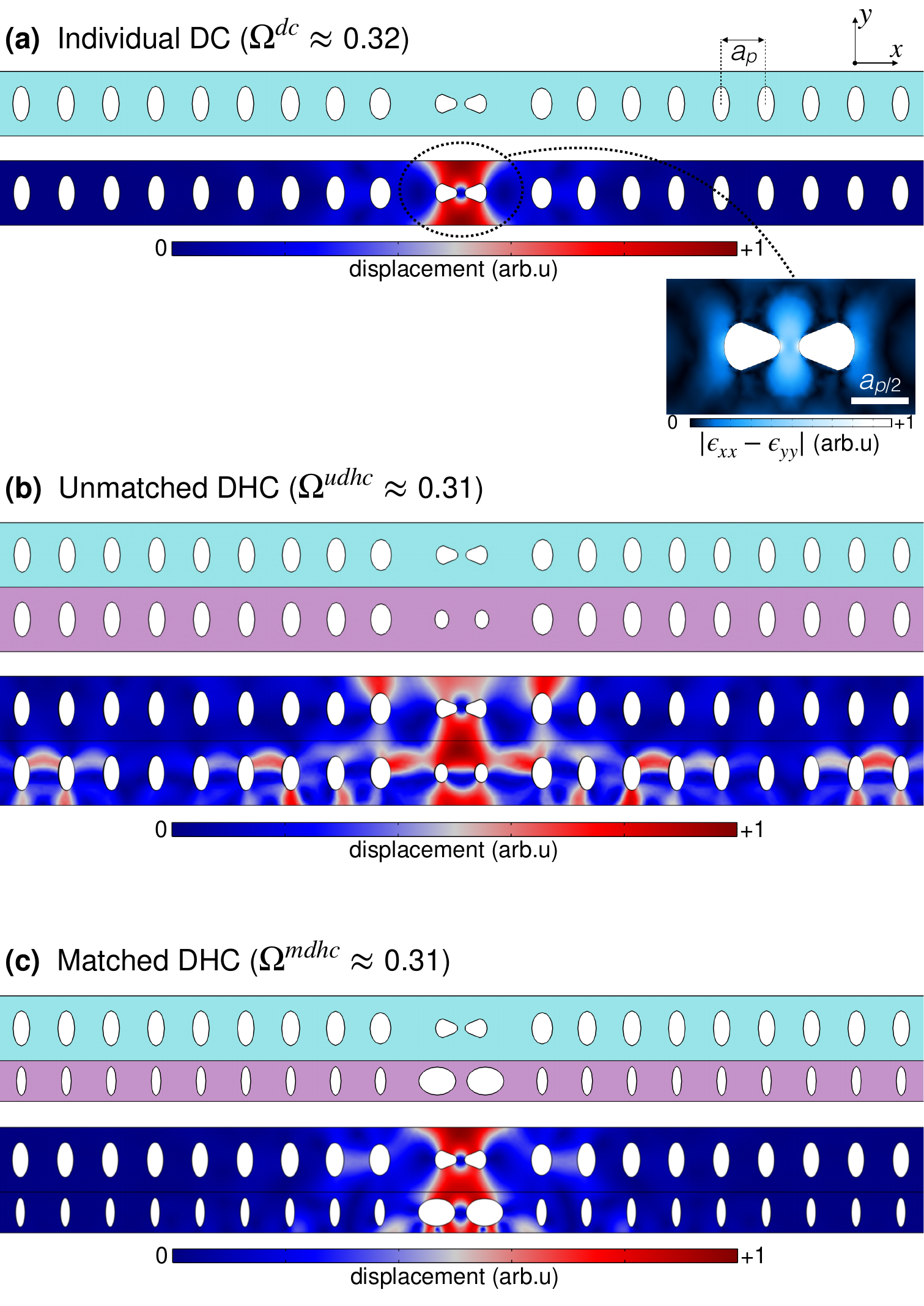}
\caption{Geometry and normalized displacement profile of the fundamental eigenmode of different cavities. (a) individual DC, with eigenmode at the reduced frequency $\Omega^{dc}\approx 0.32$. Inset: $\lvert \bm{\epsilon}_{xx}(\textbf{r})-\bm{\epsilon}_{yy}(\textbf{r}) \rvert$ profile near the concentrators. (b) unmatched DHC, with not-confined eigenmode at
$\Omega^{udhc}\approx 0.31$. (c) matched DHC, with confined eigenmode at $\Omega^{mdhc}\approx0.31$. The mode replicates, in the diamond part, the individual DC mode shown in (a).} 
\label{fig:2}
\end{figure}

\subsection{Tuning the fundamental resonant mode}
To excite efficiently the NV centers, we need to design the DHC exhibiting a resonant mode. In particular, the goal is to obtain a mode in the DHC that replicates the mode in an individual DC, thus assuring that the PC doesn't alter excessively the mode coupling with the NV centers. Here, we achieve this by optimizing the design of the defects within both the PC and the DC.

First, we describe the individual DC that will form the DHC diamond part. 
Fig. \ref{fig:2}(a) shows the DC geometry, at the center of which the unit cell periodicity is broken by the insertion of eight defects with gradually deformed holes, characterized by increasing $h_x$ and decreasing $h_y$ while moving towards the center (more details about the geometry can be found in the Appendix \ref{app:A}). The shape and the position of the two most central holes are additionally deformed into triangular concentrators \cite{schmidt2020acoustic, raniwala2022spin,kim2023diamond}, with the scope of increasing the strain and stress of confined resonant modes in the center of the cavity, where the NV centers are embedded \cite{raniwala2022spin, smith2021nitrogen}. The displacement profile of the fundamental DC resonant mode at a reduced frequency $\Omega^{dc} \approx 0.32$ is characterized by a low-displacement/high-strain between the concentrators, similarly to other cavities with comparable structures \cite{raniwala2022spin,kim2023diamond}. This is shown in the inset, reporting the difference between the $xx$ and $yy$ strain tensor components, $\lvert \bm{\epsilon}_{xx}(\textbf{r})-\bm{\epsilon}_{yy}(\textbf{r}) \rvert$.
This term contributes to the spin-mechanical coupling rate $g_{sm}^{(n)}$ between the mechanical $n$-eigenmode and the NV centers \cite{schmidt2020acoustic, raniwala2022spin}:
\begin{equation}\label{eq:g_sm}
g_{sm}^{(n)}(\textbf{r}) = \chi \frac{\bm{\epsilon}_{xx}(\textbf{r})-\bm{\epsilon}_{yy}(\textbf{r})}{\text{max}(\lvert \textbf{u}(\textbf{r}) \rvert)} x_{zpf},
\end{equation}
where $\chi=-0.85$ PHz/strain is the strain susceptibility constant for NV centers \cite{lee2016strain}, $\textbf{u}$ is the displacement profile of the $n$-eigenmode and $x_{zpf}$ its cavity zero point fluctuation. The latter can be calculated as \cite{raniwala2022spin, safavi2014optomechanical}:
\begin{equation}\label{}
x_{zpf} = \sqrt{\frac{\hbar}{2 m_{eff}\omega^{(n)}}},
\end{equation}
where $\omega^{(n)}=2 \pi v_t \Omega^{(m)}  / a_p$ is the angular frequency of the $n$-eigenmode, $\hbar$ the reduced Planck constant and $m_{eff}$ is the cavity effective mass for that mode, given by:
\begin{equation}\label{}
m_{eff}=\int d\textbf{r} {\frac{\textbf{u*}(\textbf{r})\rho(\textbf{r})\textbf{u}(\textbf{r})}{\text{max}(\lvert \textbf{u}(\textbf{r}) \rvert )^2}},
\end{equation}
$\rho(\textbf{r})$ being the mass density profile.\\
To calculate the coupling rate $g_{sm}^{dc}$ for the eigenmode in Fig. \ref{fig:2}(a), we need to fix the DC size. For a direct comparison with the cavity with the piezoelectric pad \cite{kim2023diamond}, we set a similar periodicity $a_p=270$ nm. With this choice, the two cavities present similar values for both the fundamental eigenfrequency ($\approx 15$ GHz for the DC) and the spin-mechanical coupling rate ($\lvert g_{sm}^{dc} \rvert / 2\pi \approx 20$ MHz). Although NV centers transitions frequency can be tuned by the application of a magnetic field \cite{shandilya2021optomechanical, diep2010optical}, lower frequencies ($\approx 3-5$ GHz) are more common to trigger these transitions \cite{xu2019quantum}. To obtain lower frequencies in the DC, we can scale its size by a factor $\xi$ ($a_p \rightarrow \xi a_p \Rightarrow \omega^{dc} \rightarrow \omega^{dc} / \xi$). Accordingly, the spin-mechanical coupling rate scales as $\propto 1/\xi^2$. This means that, when setting a frequency 5 GHz, as a result we get $\lvert g_{sm}^{dc} \rvert / 2\pi \approx 2$ MHz. For 3 GHz, $\lvert g_{sm}^{dc} \rvert / 2\pi \approx 1$ MHz. Thus, in a real experiment, there will be a trade-off between a high spin-mechanical coupling rate, which requires small $a_p$ and high frequencies, and the achievable frequency NV resonant frequency, which depends on the applied magnetic field.

To form the DHC, we now add the PC, assuming the continuity of the displacement between the two adjoined cavities. Figure \ref{fig:2}(b) shows the unmatched DHC made by identical diamond/AlN unit cells. Also the defects are identical, with the only exception of the concentrator holes, which are not included in the PC. The unmatched DHC presents a series of not-confined mechanical eigenmodes, among which Fig. \ref{fig:2}(b) shows the most closely resembling one to the fundamental individual DC eigenmode (Fig. \ref{fig:2}(a)). This clearly shows how, in general, the piezoelectric transducer can induce an alteration of the diamond mode and a deterioration of its features.
The mode has a reduced frequency $\Omega^{udhc}\approx 0.31$ GHz and a normalized mode-volume of $\approx 0.7$. 
The latter is calculated as the temporal average of $\frac{1}{a_p^3}\int d\textbf{r}\mathscr{U}_{mec}/\text{max}(\mathscr{U}_{mec})$, i.e. the integrated mechanical energy density (Eq.\ref{eq:energydensity}) divided by its maximal value and normalized by $a_p^3$.
Interestingly, separating the integral in the PC and DC contributions, we notice that the PC contributes for the most part of the volume ($\approx 75 \%$), indicating that the non-confinement is mostly due to the PC. This is evident looking at the eigenmode in Fig. \ref{fig:2}(b), where phonon leakage primarily occurs in the PC, thereby hindering efficient excitation of the DC. We highlight that this is not due to the particular choice of the PC defects geometry, but it is the direct consequence of the absence of bandgaps associated to the unmatched combined unit cell, as discussed above with reference to Fig. \ref{fig:1}(b).

Conversely, when using the matched double unit cell to form a DHC, a resonant mode within the bandgap can be obtained.
Fig. \ref{fig:2}(c) shows the matched DHC cavity and its well confined fundamental eigenmode at $\Omega^{mdhc} \approx 0.31$. The good confinement is reflected also in the normalized mode-volume, being $\approx 0.09$, i.e. almost one order of magnitude smaller than the one of the unmatched DHC. The mode was obtained by optimizing the design of the defects position and size in both the DC and the PC. Such a structural optimization was performed by a homemade genetic algorithm, whose details can be found in the Appendix \ref{app:C}. Regarding the robustness of the mode, in the Appendix \ref{app:D} we show  that it can tolerate modifications of approximately $\pm$10\% on the PC and DC defects sizes, without changing significantly its spin-mechanical coupling and mode-volume. This result gives to the cavity a good tolerance to possible defects introduced during an experimental fabrication process.\\
Most importantly, the displacement field inside the DC part reproduces almost perfectly the fundamental eigenmode of the individual DC in Fig. \ref{fig:2}(a). This demonstrates that through the precise matching of the phononic properties of the cavities forming the DHC, we were able to design a piezoelectric transducer that preserves the displacement field shape of the individual DC.
Having similar displacement, we expect also the coupling with the NV centers, $g_{sm}$, to be similar for the two cavities, although in the DHC the additional PC part increases $m_{eff}$. 
To calculate the $g_{sm}$ in case of the DHC, we need to generalize the expression in Eq.\ref{eq:g_sm}: $\text{max}(\lvert \textbf{u}(\textbf{r})\rvert )$ and $ x_{zpf}$ are calculated over the entire DHC, while $\chi$ is set to zero in the PC.
In this way, we are able to calculate the spin-mechanical coupling rate $\lvert g_{sm}^{mdhc}\rvert$ for the matched DHC eigenmode in Fig. \ref{fig:2}(c). The ratio between its maximal value and the previously calculated one in the individual DC is $\lvert g_{sm}^{mdhc} \rvert / \lvert g_{sm}^{dc} \rvert \approx 0.8$. As expected, the ratio is close to 1, demonstrating that the presence of the optimized PC has only a moderate effect on the spin-mechanical coupling in the DHC.\\
\begin{figure}[ht]
\centering
\includegraphics[width=0.49\textwidth]{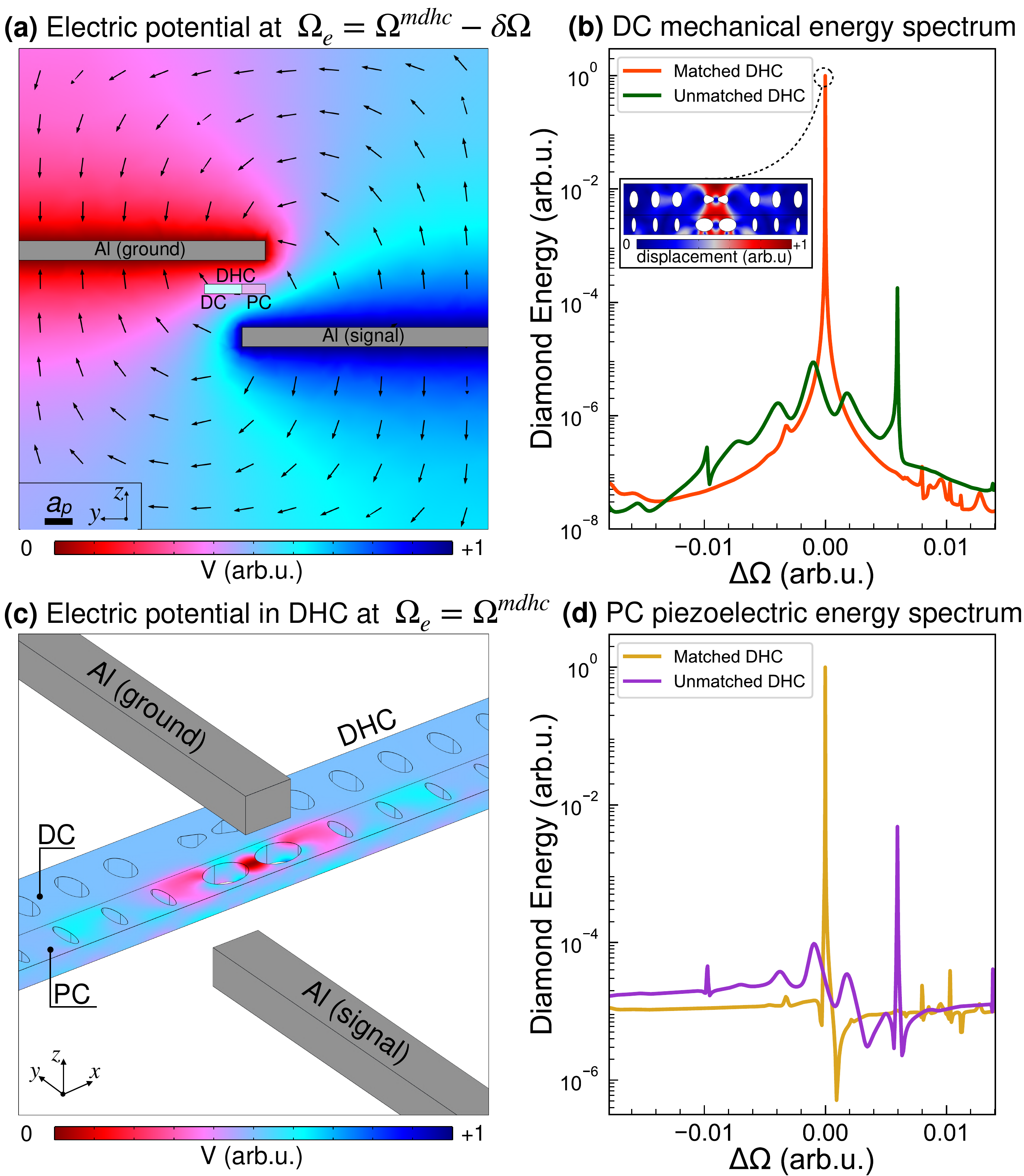}
\caption{Non-contact microwave photon-to-phonon excitation via piezoelectric effect in the DHC. (a) electric potential profile in the $y$-$z$ plane at $\Omega_e=\Omega^{mdhc}-\delta \Omega$, with $\delta \Omega\approx 2 \times 10^{-4}$. (b) mechanical energy in the central part of the DC as a function of $\Delta \Omega = \Omega_e - \Omega^{mdhc}$ for the unmatched DHC (green curve) and matched DHC (orange curve). Inset: displacement profile for the excited resonant mode at $\Delta \Omega_e=0$ in the matched DHC. (c) electric potential profile inside the matched DHC at $\Delta \Omega =0$. (d) piezoelectric energy $U_{pe}$ in the PC as a function of $\Delta \Omega$ for the unmatched DHC (purple curve) and matched DHC (yellow curve).
}
\label{fig:3}
\end{figure}
\subsection{Excitation by non-contact electrodes}
The main advantage of diamond/piezoelectric hybrid devices is the possibility to excite  NV centers embedded in diamond via the inverse piezoelectric effect. Here, to activate the DHC mechanical modes, we employ microwave non-contact electrodes positioned, respectively, below and above the DHC, following the approach proposed by H{\"o}nl \textit{et al}. \cite{honl2022microwave}.

The electrodes generate an electric field, triggering the PC piezoelectric response. The mechanical energy produced in the PC actives acoustical modes in the entire DHC, including the DC. The PC, therefore, acts as a photon-to-phonon transducer. 
The voltage between the electrodes is sinusoidally modulated at reduced frequency $\Omega_e$. Figure \ref{fig:3}(a) shows the electric potential around the matched DHC in the $y$-$z$ plane, generated by the electrodes at $\Omega_e= \Omega^{mdhc} - \delta     \Omega$, i.e. close to the matched DHC resonance mode ($\delta   \Omega \approx 2 \times 10^{-4}$).
The arrows indicate the in-plane electric field direction, originating from the bottom electrode (signal) and directed toward the top electrode (ground). By tuning $\Omega_e$, we can observe the DHC acoustical response at various frequencies, allowing to derive its spectrum.\\
Fig. \ref{fig:3}(b) shows the spectrum of the mechanical energy stored in the central part of the DC for the unmatched DHC (green curve) and the matched DHC (orange curve) as a function of $\Delta \Omega = \Omega_e - \Omega^{mdhc} $.
The energy is calculated using Eq.\ref{eq:mecenergy}, but this time the spatial integral is performed over the central part of the DC ($x \leq \lvert 0.9a_p \rvert$), since we are only interested in confined modes. The spectrum spans over a frequency range corresponding to the bandgap of the matched DHC, as confined modes are only possible within this range. The matched DHC exhibits a prominent peak at the frequency of its resonant mode ($\Delta \Omega = 0$, $\Omega_e=\Omega^{mdhc}$), which is several orders of magnitude higher than any peak observed in the unmatched DHC. In the inset, we present the displacement profile of the matched DHC mode excited by the electrodes, which precisely matches the eigenmode (Fig. \ref{fig:2}(c)). Consequently, employing non-contact electrodes proves to be a suitable strategy to excite the fundamental mode of the matched DHC, and thus trigger the embedded NV centers.

The efficiency of the exaction is determined by the microwave-to-acoustic conversion via the piezoelectric transducer. Therefore, we need to demonstrate that the choice of the PC as a transducer not only doesn't affect the feature of the fundamental diamond mode, but is also capable of providing an efficient excitation. A high conversion rate within the PC, indeed, implies a more efficient control of the mechanical excitation in the DHC and thus of the spin transitions at a given voltage between the electrodes. This conversion rate can be quantitatively evaluated by the piezoelectric coupling \cite{arrangoiz2018coupling}. To calculate it for the DHC, we start by taking the total piezoelectric energy in the PC, given by the integral of the stress field \textbf{T} and the electric field \textbf{E}:
\begin{equation}\label{}
    U_{pe}=\frac{1}{2} \int_{PC} d\textbf{r}(\textbf{T}\cdot\textbf{d}^T\cdot\textbf{E} + \textbf{E} \cdot\textbf{d} \cdot \textbf{T}),
\end{equation}
where \textbf{d} is the piezoelectric tensor in the PC. High values of $U_{pe}$ are then expected for configurations presenting a significant overlap between \textbf{T} and \textbf{E}. Figure \ref{fig:3}(c) shows the electric potential within the matched DHC at $\Omega_e=\Omega^{mdhc}$ ($\Delta \Omega=0$), revealing a well confined peak at the center of the PC. Consequently, the optimized patterning in the PC leads to a tight confinement of both the stress and the electric field in the central part of the matched DHC cavity. Figure \ref{fig:3}(d) shows the spectrum of $U_{pe}$ for the unmatched DHC (purple curve) and the matched DHC (yellow curve). As expected, we observe a high peak for the matched DHC at $\Delta \Omega=0$ ($\Omega_e=\Omega^{mdhc}$), where $\textbf{T}$ and $\textbf{E}$ are strongly confined and overlap significantly. This peak is order of magnitudes higher than any other peak for the unmatched DHC, whose photons and phonons are poorly confined.\\ 
However, it is important to note that $U_{pe}$ depends on the voltage magnitude between the electrodes, so its absolute value alone doesn't provide explicit information about the piezoelectric conversion efficiency. To assess the latter, it is necessary to normalize both \textbf{T} and \textbf{E}.
To this end, \textbf{T} and \textbf{E} can be decomposed as a sum of the normalized eigenmodes $\textbf{T}^{(n)}$ and $\textbf{E}^{(m)}$. Then, the piezoelectric coupling $g_{pe}^{n-m}$ between a mechanical $n$-mode and an electrical $m$-mode is \cite{zou2016cavity}:
\begin{equation}\label{}
g_{pe}^{n-m}=\frac{1}{2 \hbar} \int_{PC} d\textbf{r}(\textbf{T}^{(n)}\cdot\textbf{d}^T\cdot\textbf{E}^{(m)} + \textbf{E}^{(m)} \cdot\textbf{d} \cdot \textbf{T}^{(n)}),
\end{equation}
whose value solely depends on the profile of the involved fields.
In the case of the DHC and electrodes system, the electric field \textbf{E} is generated by two sources, namely the electrodes and the PC, with the second one due to the piezoelectric effect. \textbf{T} is instead determined only by the stress within the PC part of DHC modes. When $\Omega_e$ coincides with the DHC reduced resonant frequency $\Omega^{(\tilde{n})}$ of the $\tilde{n}$-eigenmode (with angular frequency $\omega^{(\tilde{n})}=2 \pi v_t \Omega^{(\tilde{n})}  / a_p$), both \textbf{T} and \textbf{E} oscillate at $\Omega_e = \Omega^{(\tilde{n})}$. In other words, the decomposition of each field occurs only on the $\tilde{n}$-eigenmode: $\textbf{T} \propto  \textbf{T}^{(\tilde{n})}$ and $\textbf{E} \propto \textbf{E}^{(\tilde{n})}$. For the proportionality factor, we opt for the square root of the ratio between the energy carried by the quantum of the field (phonon or photon) and the total field energy, ensuring the normalization of eigenmodes:
\begin{eqnarray}
	\textbf{T}^{(\tilde{n})} & = & \sqrt{\frac{ \hbar \omega^{(\tilde{n})}}{U_{mec}}}\textbf{T},
	\nonumber\\
    \label{eq:normalization}\\
   \textbf{E}^{(\tilde{n})} & = &   \sqrt{\frac{\hbar \omega^{(\tilde{n})}}{U_{em}}}\textbf{E}.
	\nonumber
\end{eqnarray}
The total mechanical energy of the system, $U_{mec}$, is calculated as in Eq.\ref{eq:mecenergy}, with the integral spanning along the entire DHC. The total energy generated by the electric field, $U_{em}$, is instead given by
\begin{equation}
U_{em} = \frac{1}{2\tau^{(\tilde{n})}} \int_0^{\tau^{(\tilde{n})}} d\text{t} \int d\textbf{r} \varepsilon \lvert \textbf{E} \rvert^2, 
\end{equation}
where $\tau^{(\tilde{n})}=2 \pi /\omega^{(\tilde{n})}$and the spatial integral extends over the PC, the DC and the surrounding vacuum, and the permittivity $\varepsilon$ changes accordingly.\\
As for the spin-mechanical coupling rate, also the piezoelectric coupling depends on the dimension of the cavity. Therefore, we set again $a_p=270$ nm for a direct comparison with the cavity where the microwave-to-phonon conversion is provided by the piezoelectric pad \cite{kim2023diamond}. With this choice, we obtain a value of the piezoelectric coupling rate of $ \lvert g_{pe}^{mdhc} \rvert / 2\pi \approx 2$ GHz, four order of magnitude higher than in the cavity with the pad \cite{kim2023diamond}. When tuning the DHC resonant mode at lower NV color centers frequencies ($ \approx 3-5$ GHz \cite{xu2019quantum}), we still obtain the high value of $ \lvert g_{pe}^{mdhc} \rvert / 2\pi \approx 0.4-0.7$ GHz.
This result shows the effectiveness of using a PC as the transducer.
Such high values, indeed, can be rationalized by considering the high confinement of the mechanical stress and electric field provided by the nanocavity. The small mode-volume of the PC resonant mode leads to a significant overlap between the two fields, thus causing a large enhancement in the piezoelectric coupling rate.\\
\section{Conclusion}
In this work, we theoretically introduced the design of a quasi-one dimensional phononic DHC formed by two adjoined cavities of diamond and AlN. Within the DHC, the DC has the potential to incorporate NV color centers, while the PC assures an efficient control over mechanical excitation of the entire structure. We demonstrated that the choice of a PC as a piezoelectric transducer presents two ideal characteristics. First, the PC can be designed so that the diamond fundamental mode remains almost unaltered, ensuring a high coupling with the NV centers. Second, the PC presents a high piezoelectric coupling rate, demonstrating its effectiveness as a transducer.

To obtain the design of the DHC, we first tuned the phonon dispersion of the individual diamond/AlN unit cells to obtain a bandgap in the phonon dispersion of the combined hybrid double cell. This, in turn, allowed us to design a matched DHC, presenting a well-confined mode with frequency falling within the bandgap. We tailored this mode using a genetic algorithm optimization on the DHC defects, obtaining as a result a fundamental mode that closely replicates the profile of the fundamental mode in the individual DC. The similarity of the modes leads to a spin-mechanical coupling rate ratio between the two cavities of $\approx 0.8$. This shows that the presence of the PC only has a moderate effect on the spin-mechanical coupling rate in diamond. In discussing such a coupling, we also pointed out that, in a real experiment, a trade-off must be found between a high spin-mechanical coupling rate, which can be enhanced by shirking the cavity size and thus increasing its resonant frequency, and the obtainable NV resonant frequency, which can be tuned by an external magnetic field.\\
In the AlN part of the cavity, instead, we predicted a piezoelectric coupling rate four order of magnitude higher than for the cavity excited by a piezoelectric pad \cite{kim2023diamond}, when the two cavities have similar size.
We attributed such a high value to the strong overlap between the confined stress and electric fields assured by the use of a phononic cavity as the piezoelectric transducer.\\
The resonant mode has also proven to be resistant to possible variations in the size of cavity defects, which can be introduced during the fabrication process. The mode maintains a good performance in terms of spin-mechanical coupling and mode-volume within a range of approximately $\pm 10$\% on defect size in PC and DC. Still regarding the  fabrication process, we can make a comment about the assumption of perfect mechanical contact between diamond and AlN. In this respect, the deposition of AlN on the diamond \cite{shirato2020high, ishihara2002preparation} and other materials \cite{bengtsson1996applications, iqbal2018reactive} is performed by sputtering and subsequent annealing at high temperatures. This technique ensures an optimal adhesion between the two materials, justifying our assumption.

Finally, we presented a scheme in which non-contact electrodes, positioned respectively above and below the cavity, successfully excite the resonant mode of the entire DHC, thus triggering the color centers transitions in diamond.

The color centers states can thus be detected by optical techniques \cite{gruber1997scanning, macquarrie2013mechanical, hausmann2012integrated}. In this process, the presence of the PC adjoined to the DC is expected to have a small impact, since photons emitted by NV centers will stay confined in the DC \cite{kim2023diamond}, being the diamond refractive index higher than the AlN one.
Nevertheless, a natural evolution of our system would be a diamond/piezoelectric double hybrid optomechanical cavity that exploits the advantages demonstrated here by using the two cavities. The confinement of photons due to the optomechanical cavity , indeed, will enhance the read-out of the diamond spin qubit state.

\begin{acknowledgments}
This work was supported by Japan Science and Technology Agency Moonshot R\&D grant (JPMJMS2062), by the JSPS KAKENHI (Grant Number 21H04635), by the JSPS KAKENHI Research Activity Start-up (Grant Number JP23K19196) and by the JSPS KAKENHI (Grant Number JP23KF0203).
\end{acknowledgments}

\appendix
\section{Geometry and material properties}
\label{app:A}
The principal crystallographic axes of all the unit cells of this work are assumed to be along the Cartesian coordinates. This means that, for diamond, the three equivalent [100], [010] and [001] crystallographic directions are along, respectively, the $x$, $y$ and $z$ coordinates. For AlN, the out-of-plane lattice axis is taken along the $z$-coordinate and in-plane equivalent lattice axes along, respectively, $x$ and $y$ coordinates. In this way, diamond/AlN elastic constants and AlN piezoelectric coupling constants take the values reported in Table \ref{table:materialproperties}, written in Voigt notation.
\begin{table}[htb]
\centering
  \caption{Material properties}
  \label{table:materialproperties}
  \begin{tabular}{ccc}
  \hline
  \hline
    Parameter & Diamond & AlN \\
    \hline
    $\varepsilon/\varepsilon_0$  & 5.9 & 9      \\
    $\rho$ (kg/m$^3$) & 3515 & 3300   \\
    $C_{11}$ (GPa)  & 1076 & 410   \\
    $C_{12}$ (GPa)  & 125 & 149     \\
    $C_{13}$ (GPa)  & 125 & 99     \\
    $C_{22}$ (GPa)  & 1076 & 410  \\
    $C_{23}$ (GPa)  & 125 & 99    \\
    $C_{33}$ (GPa)  & 1076 & 389   \\
    $C_{44}$ (GPa) & 578 & 125   \\
    $C_{55}$ (GPa) & 578 & 125    \\
    $C_{66}$ (GPa) & 578 & 130.5   \\  
    \textbf{d}$_{15}$ (C/m$^2$)  & / & -0.48   \\
    \textbf{d}$_{24}$ (C/m$^2$)  & / & -0.48   \\
    \textbf{d}$_{31}$ (C/m$^2$)  & / & -0.58   \\
    \textbf{d}$_{32}$ (C/m$^2$)  & / & -0.58   \\
    \textbf{d}$_{33}$ (C/m$^2$)  & / & 1.55   \\
    \hline
    \hline
  \end{tabular}
\end{table}\\

From a geometrical point of view, all unit cells present a length $a_p$ along the $x$-axis and a thickness $t=0.37 a_p$. The diamond unit cell and the identical (unmatched) AlN unit cell present a rectangular cross section with width $w=1.43 a_p$ and a patterned elliptical hole at its center, with axis $h_x=0.37 a_p$ and $h_y=0.78 a_p$ (see Fig. \ref{fig:1}(a)). The optimized (matched) AlN unit cell is obtained by altering the previous parameters to $w'=0.9 a_p$, $h_x'=0.22 a_p$ and $h_y'=0.65 a_p$ (see Fig. \ref{fig:1}(c)).
\begin{figure}[htb]
\centering
\includegraphics[width=0.45\textwidth]{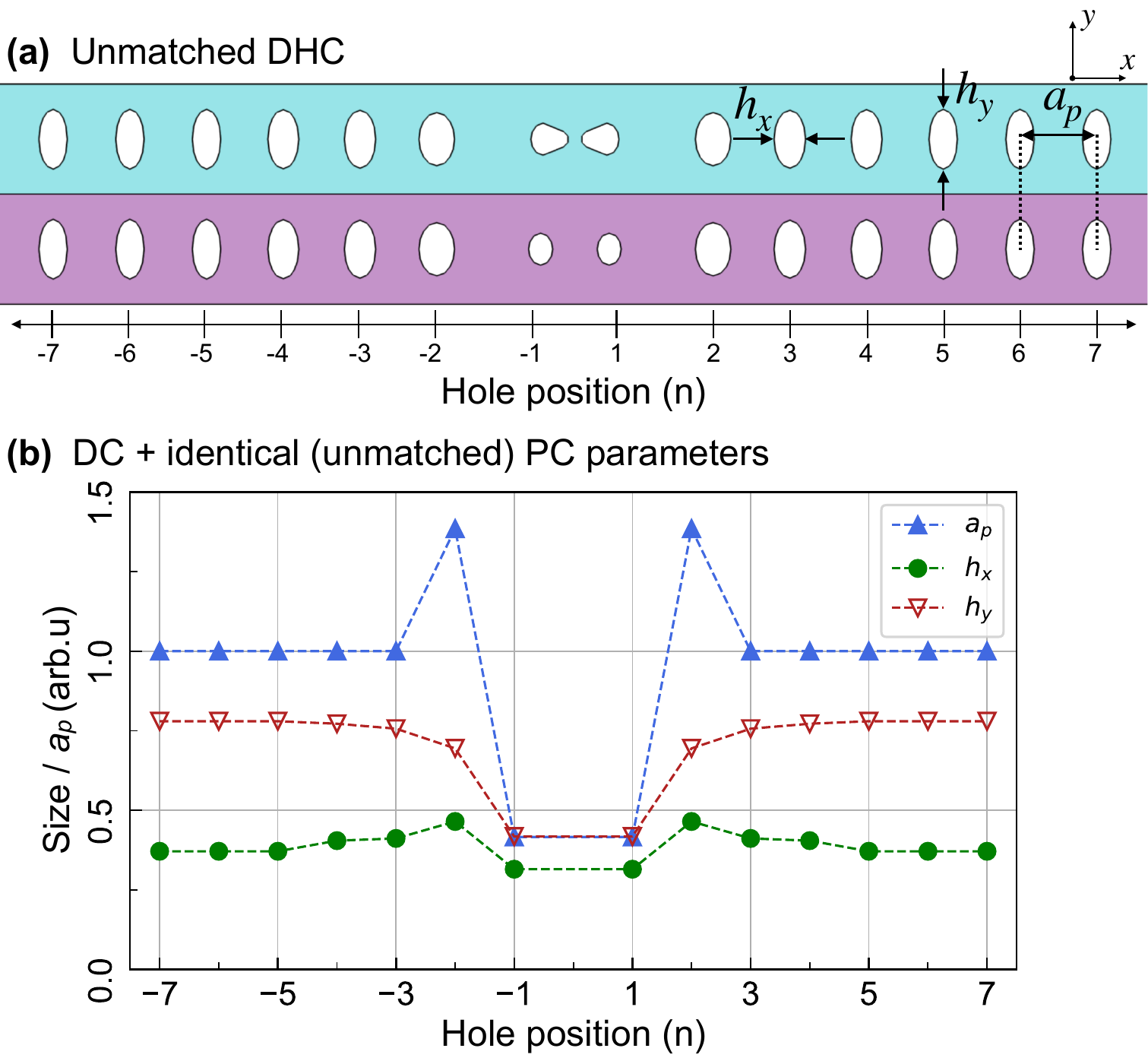}
\caption{Central part of the unmatched DHC (a) and behaviour of the different parameters of the DC and identical (unmatched) PC as a function of the hole position (b).}\label{fig:a4}
\end{figure}
The unmatched and the matched unit cells are used to form, respectively, the unmatched and matched DHCs. Both cavities are made with 46 holes, of which 38 are given by the repetition of the corresponding unit cells described above and the 8 central ones are deformations (defects) of the cells. The deformation is given by a modification of the periodicity and of the axes of the elliptical holes.
Fig. \ref{fig:a4}(a) shows the central part of the unmatched DHC. The holes are numbered with the index n ($-7\le \text{n} \le 7$), thus allowing to describe their parameters as a function of their position. Figure \ref{fig:a4}(b) shows the behaviour of $a_p$, $h_x$ and $h_y$ within the DC and the identical PC forming the unmatched DHC. In addition to the elliptical holes, the DC also presents two triangular holes that superimpose those at n=$\pm 1$. The tip of each triangular hole points to the center of the cavity and is rounded to make it suitable for fabrication. The distance between the two triangles is $\approx 0.16 a_p$.

\begin{figure}[htb]
\centering
\includegraphics[width=0.45\textwidth]{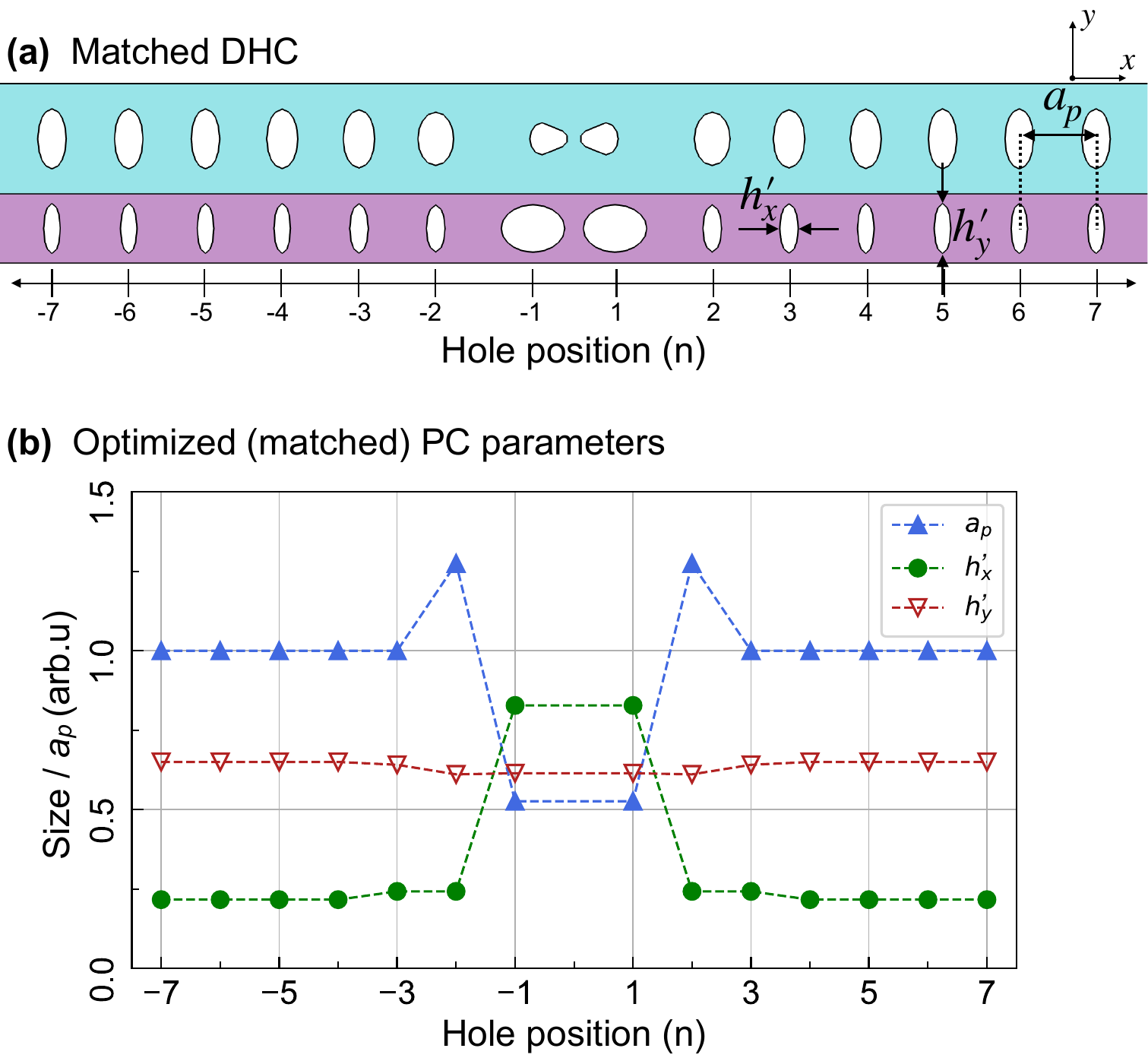}
\caption{Central part of the matched DHC (a) and behaviour of the different parameters of the optimized (matched) PC as a function of the hole position (b).}\label{fig:a5}
\end{figure}
Figures \ref{fig:a5}(a) and \ref{fig:a5}(b), instead, displays respectively the central part of the matched DHC and the behaviour of $a_p$, $h_x'$ and $h_y'$ for the PC as a function of n. Also in this case, holes with  $\lvert \text{n} \rvert \le 4$ present a deformation from the original unit cell. The optimization of these parameters was performed via genetic algorithm in combination with finite element methods (see Appendix \ref{app:C}).

\section{Finite element simulations}
\label{app:B}
\subsection*{Phonon dispersion relations}
Finite element simulations on single and double unit cells were performed to obtain their phonon dispersion curves. Free boundary conditions were used at the cell/vacuum boundaries, while Floquet periodic boundary conditions were imposed at the cells external faces perpendicular to the $x$-axis. To obtain the phonon dispersion, we calculated the cells mechanical eigenmodes as a function of the wavevector $K$. Due to the periodicity of the system and the folding of the branches, $K$ only spans the 1-d Brillouin zone between $-\pi / a_p$ and $ + \pi / a_p$.
Since we are interested only in $z$-symmetric modes, we simulated only half of the cell, accounting only for half of its thickness $t$. A symmetric boundary condition was applied on the resulting exposed surface. All the phonon dispersion curves in the main text are $z$-symmetric. Furthermore, single cells are also divided in half along the $y$-direction. In this way, we can apply to the resulting exposed surface $y$-symmetric and $y$-antisymmetric boundary conditions and obtain the corresponding symmetric and antisymmetric modes. In the double cells, instead, due to the presence of different materials, the periodicity along the $y$-direction is broken, so no $y$-symmetric or $y$-antisymmetric condition can be applied.

\subsection*{Phononic cavities}
Finite element simulations on the individual DC, the unmatched DHC and the matched DHC were performed to obtain their mechanical resonant modes via eigenmode study. Then, we retrieved the profiles of the displacement, strain, stress, mode-volume and spin-mechanical coupling associated to these modes.
For the two DHCs, simulations in the frequency domain were also performed to calculate the acoustic/piezoelectric response of the cavities under excitation by an electric field generated from the two top/bottom electrodes. The simulations are inherently multiphysics, taking into account the reciprocal feedback of the mechanical and the piezoelectric responses of the system in the frequency domain. In particular, we calculated the energy in the central part of the DC and the piezoelectric energy in the PC, as well as the piezoelectric coupling constant. The electric field frequency was swept to derive the spectral response of the cavities. Spectra in Fig. \ref{fig:3} of the main text are calculated with a $\Omega-$step of $\approx 2 \times10^{-5}$, but around the peaks the step was decreased to $\approx 2.5 \times10^{-6} $ for a finer analysis.

For the mechanical part of the study, a free boundary condition was applied to all cavity/air interfaces. At the end of the cavity, a perfectly matched layer boundary condition was applied to avoid artificial internal reflections of the excited mechanical modes. Electrodes were not included in the mechanical study, since they are not in mechanical contact with the cavity.\\
For the piezoelectric study, the two electrodes take the role of, respectively, the terminal and the ground, between which a voltage difference is fixed. Moreover, we used an air domain surrounding the cavity, in order to take into account the electric field generated by the electrodes. A zero charge condition is applied at the external boundaries of the air domain.\\
For the eigenmodes simulations, we simulated both the entire cavities and their halves, applying a symmetric condition at the $y$-$z$ plane cutting in a half the cavities.
For the frequency domain simulations, due to the higher computational cost, we accounted only for half of the cavity and applied a symmetric boundary condition.

\begin{figure}[htb]
\centering
\includegraphics[width=0.375\textwidth]{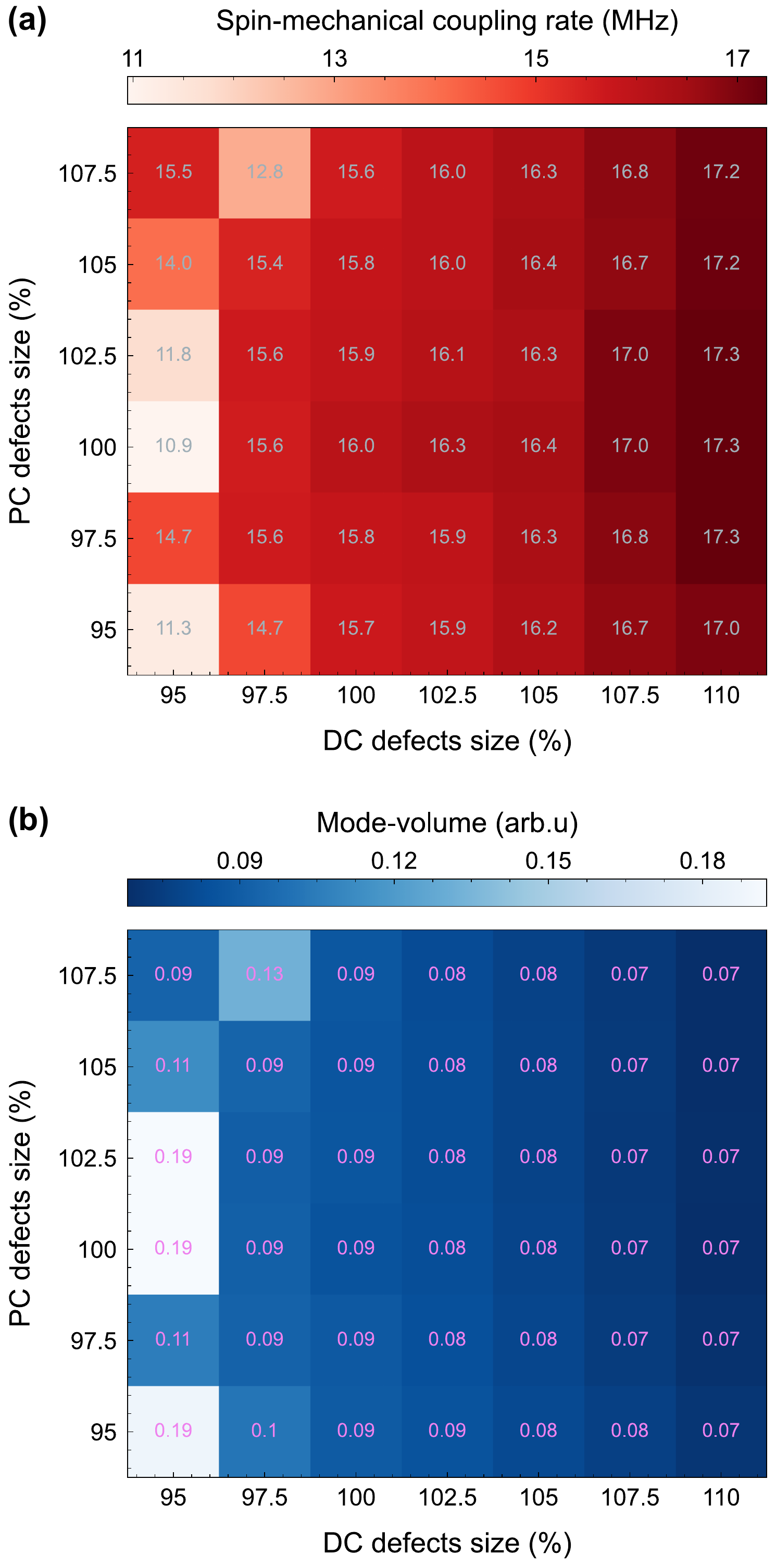}
\caption{Spin-mechanical coupling rate (a) and the mode-volume (b) of the matched DHC resonant mode as a function of the size of the defects in the DC and the PC.
}\label{fig:a6}
\end{figure}

\section{Genetic algorithm optimization}
\label{app:C}
The optimization of the matched DHC defect parameters was obtained via a homemade genetic algorithm. The algorithm varies the defects parameters $a_p$, $h_x$ ($h_x'$), $h_y$ ($h_y'$) and automatically incorporates them into the finite element simulation. The results of each simulation are then fed back into the genetic algorithm that determines the evolution of the parameters. The genetic algorithm code was implemented on a binary gene encoding scheme, incorporating a dynamic penalty function to handle constraints on the parameters. The constrains were imposed to keep the holes of the cavity within the typical dimensions achievable in a clean room fabrication process.
Additionally, we introduced a sharing function to ensure extensive exploration of the search parameters space.

Two figures of merit were optimized to select the final parameters of the matched DHC, namely the spin-mechanical coupling rate $g_{sm}$ and the quality factor $Q$ of the fundamental mechanical mode. We observed that $g_{sm}$ is largely influenced by the most central defects (n$=\pm1$), while $Q$ is more sensible to the other three defects (n=$\pm 2,\pm 3, \pm 4$). To take into account both these two figurs of merit, we opted for an iterative optimization process. Initially, we concentrated on enhancing $g_{sm}$ by adjusting the most central defects (n$=\pm1$) of both the PC and the DC. The algorithm settings for this first step were a ``Boltzmann'' selection method, a population of 20 individuals and 20 generations. Then, we carried out a second optimization step focusing only on the three other defects to improve $Q$, again both for the DC and the PC. In this optimization step, we kept the same selection method as before, but executed two different runs, the first one with a population of 30 individuals and 20 generations, the second one with a population of 25 and 20 generations. Finally, we conducted a third run to maximize $g_{sm}$, once again focusing on the central defects. This third and final optimization step used an ``harmonic'' selection method, with a population of 30 individuals and 50 generations.

For all the optimization processes the evolution parameters were fixed as: crossover probability =80$\%$, elitism probability=10$\%$, mutation probability=8$\%$.

The final quality factor obtained via genetic algorithm for the fundamental mode in the matched DHC is $ \approx 5 \times 10^7$. Nevertheless, this was obtained with perfectly matched layer conditions at the edges of the cavity, and doesn't take into account thermal noise phenomena that can severely affect its value. Therefore, although it is useful to maximise $Q$ so as to have the best possible structure parameters, its value is not reported in the main text.

\section{Effect of imperfections on the matched DHC resonant mode}
\label{app:D}
To test the tolerance of the matched DHC to imperfections that can occur during a fabrication process, we calculated the resonant mode at different defects sizes in the DC and the PC. Figure \ref{fig:a6} reports the spin-mechanical coupling rate $g_{sm}/2\pi$ (a) and the normalized mode-volume (b) of the matched DHC resonant mode as a function of the defects size. The sizes are normalized with the ones of a cavity having $a_p=270$ nm, as in the main text. The resonant mode is very robust, tolerating defects changes of $\approx \pm$ 10\%, while keeping reasonable values of the two figures of merit.\\
When the period is set to approximately 800 nm, corresponding to the excitation frequency of NV centers, the defects in the DHC have the size of hundreds of nm, thus a 10\% tolerance means a required fabrication precision of tens of nm, which is feasible with today fabrication techniques.

Notice that the best performances for both the spin-mechanical coupling rate and the mode-volume are for large values of the DC defects. Nevertheless, DC defects of that size can create fragile thin edges in the cavity, that would probably break during the fabrication process. For this reason, we decided to keep the DHC in the main text at the center of the tolerance range, assuring a more reliable cavity that can tolerate larger defects.

\bibliography{apssamp}

\end{document}